\documentclass[aps,prb,twocolumn,showpacs,superscriptaddress,email]{revtex4-1}  
\usepackage[pdftex]{graphicx}  
\usepackage{dcolumn}   
\usepackage{bm}        
\usepackage{amssymb}   

\usepackage{amsmath}
\usepackage[latin1]{inputenc}

\usepackage{color}

\newcommand{\ket}[1]{\left| #1\right\rangle}

\newcommand\Tr{\mathrm{Tr}}

\begin{document}

\title{Exploring qubit-qubit entanglement mediated by one-dimensional plasmonic nanowaveguides.}

\author{A. Gonzalez-Tudela}
\affiliation{Departamento de F\'isica Te\'orica de la Materia
Condensada, Universidad Aut\'onoma de Madrid, Madrid 28049, Spain}

\author{D. Martin-Cano}
\affiliation{Departamento de F\'isica Te\'orica de la Materia
Condensada, Universidad Aut\'onoma de Madrid, Madrid 28049, Spain}

\author{E. Moreno}
\affiliation{Departamento de F\'isica Te\'orica de la Materia
Condensada, Universidad Aut\'onoma de Madrid, Madrid 28049, Spain}

\author{L. Martin-Moreno}
\affiliation{Instituto de Ciencia de Materiales de Aragon (ICMA)
and Departamento de Fisica de la Materia Condensada,
CSIC-Universidad de Zaragoza, E-50009 Zaragoza, Spain}

\author{F.J. Garcia-Vidal}
\affiliation{Departamento de F\'isica Te\'orica de la Materia
Condensada, Universidad Aut\'onoma de Madrid, Madrid 28049, Spain}

\author{C. Tejedor}
\email[Corresponding author: ]{carlos.tejedor@uam.es}
\affiliation{Departamento de F\'isica Te\'orica de la Materia
Condensada, Universidad Aut\'onoma de Madrid, Madrid 28049, Spain}

\begin{abstract}
We exploit the qubit-qubit coupling induced by plasmon-polariton modes in a one-dimensional nanowaveguide to obtain various two-qubit entanglement situations. Firstly, we observe three phenomena occurring when preparing the initial state of the system and leaving it freely relax: spontaneous formation of entanglement, sudden birth and revival. Then, we show that plugging a laser to each of the qubit, the system arrives to an steady state which, depending on their inter-qubit distance, can also be entangled. For this situation, we also characterize the quantum state of the system showing the entanglement-purity- diagram typical for two-qubit systems.\end{abstract}
\date{\today}
\maketitle














\section{Introduction\label{sec:intro}}

Entanglement is one of the most striking features of quantum mechanics. From the fundamental point of view, the non-separability of systems in product states of their components is a very interesting phenomenon as it has no classical analogue. From the practical point of view, entanglement is a key element of quantum cryptography, quantum teleportation, and other two-qubit quantum operations \cite{nielsen,haroche}. Although it was first studied and exploited for atomic-like systems, due to the advances in semiconductor physics and their promising scalability in fabrication important progress have been made in this direction and short-range entanglement between spin or charge degrees of freedom in quantum dots, (QDs), nanotubes, or molecules \cite{makhlin,kouwenhoven07,defranceschi} has already been observed. However, in order to get some long-distance entanglement one need some kind of interaction mediating between the qubits is needed.
Usually photons\cite{majer07a,imamoglu99a,laucht10a} have played this role, although there have also been some other proposals, for instance using  NV centers in diamond \cite{dutt07} or more recent theoretical proposals to use one-dimensional plasmon-polariton (hereafter simply labeled as plasmons) nanowaveguides to achieve steady-state entangled two-qubit states\cite{dzsotjan10a,gonzaleztudela11a}.

\begin{figure}[htb]%
\includegraphics*[width=\columnwidth]{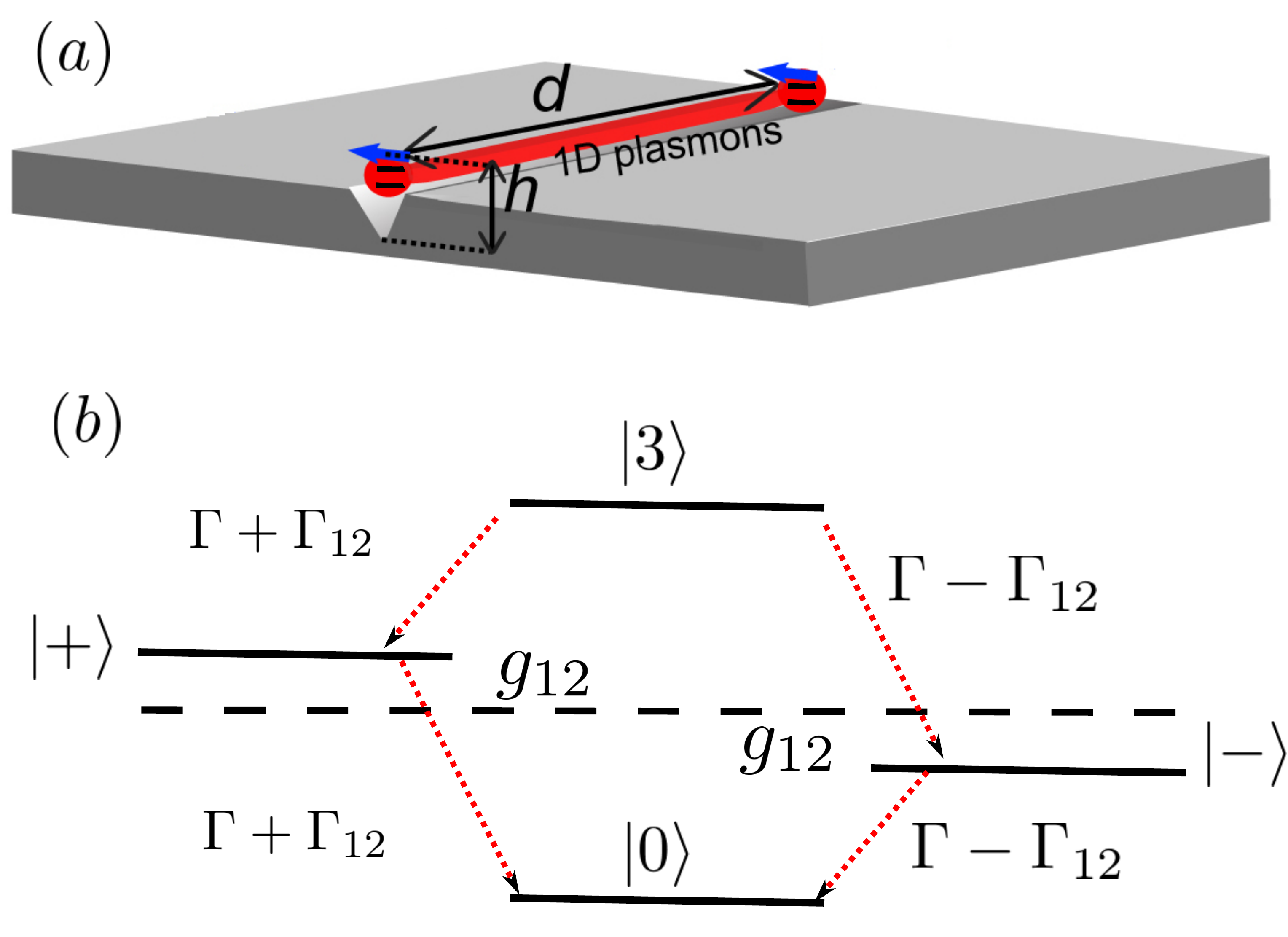}
\caption{(Color online) (a) Two qubits interacting with a plasmonic waveguide, in this case a channel waveguide. (b) Scheme of levels, couplings, and decays in the particular case where $\omega_{1}=\omega_{2}=\omega_0$ and $\gamma_{11}=\gamma_{22}=\gamma$.
}
\label{fig1}
\end{figure}

Surface plasmon-polaritons(SPP) are confined electromagnetic modes occurring in metal-dielectric interfaces\cite{raether88}. In the case of channel plasmonic waveguides(PWs) like the one of Fig. \ref{fig1}, which is the case that we will be concerned with here, these modes have one-dimensional behaviour and have been proposed as useful elements for the future generation of photonic circuits\cite{Ebbesen08}. The interest in using them for quantum optics applications\cite{chang06a} rose considerably when the coupling between a quantum emitter and a single plasmon mode was observed\cite{akimov07a}.  These works show that the $\beta$-factor, which measures the fraction of the emitted radiation that is captured by the propagating mode, can be close to $1$ due to the subwavelength nature of the plasmon field. Extension of this formalism to the case of two emitters placed close to this PWs it was also shown\cite{martincano10a} that one could get a modulable coupling between them, as a function of the inter-qubit distance,  which could be used for generating entanglement between them\cite{gonzaleztudela11a}.

The paper is organised as follows: in section \ref{sec:theor} we describe the theoretical framework that we will use for our calculations,i.e. the master equation formalism; in section \ref{sec:spontaneous} we show the results  where we consider that our system is initialized at some given state and then let it freely relax. Finally  we consider a situation where the qubits are coherently pumped in section \ref{sec:steady}, and conclude afterwards in section \ref{sec:conc}.

\section{Theoretical framework.\label{sec:theor}}

From the quantum optical point of view, we are going to see our system as two qubits, which are interacting with the plasmonic field in the PW, characterized by the continuum of bosonic modes. When the coupling between the qubits and the continuum of modes is weak, then one can safely perform a Born-Markov approximation, and trace out over the degrees of freedom of the SPP. This approximation leads to he dynamics of the density matrix $\rho$ for two qubits  described by a master equation\cite{dzsotjan10a,ficek02a} as follows:

\begin{equation}
\label{masterequation}
\partial_t \rho=i[\rho,H]+\sum_{i,j=1,2}\frac{\gamma_{ij}}{2}(2 \sigma_i \rho \sigma_j^\dagger- \sigma_i^\dagger\sigma_j\rho-\rho\sigma_i^\dagger\sigma_j)
\end{equation}
where $\sigma_i^\dagger,\sigma_i$ are the raising and lowering operators for each qubit, and where $H$, taking two qubits with characteristic frequency $\omega_i$ is given by:
\begin{equation}
\label{hamilt}
H=\hbar \sum_{i=1,2}\big(\omega_i \sigma_i^\dagger\sigma_i+ g_{12}(\sigma_1^\dagger\sigma_2+\sigma_2^\dagger\sigma_1)\big).
\end{equation}

The SPP play a double role: in the coherent part of the dynamics an effective interaction between the two quits, also present in cavity QED, is provided by the exchange of virtual plasmons:

\begin{equation}
\label{lambshift}
g_{12}=\frac{1}{\pi\epsilon_0 \hbar}\mathcal{P}\int_0^\infty \frac{\omega^2 Im[\mu_1^{*} {\bf G}(\omega,{\bf r}_1,{\bf r}_2)\mu_2]}{c^2(\omega-\omega_0)}d\omega,
\end{equation}

where $\bf{G}(\bf{r}_1,\bf{r}_2)$ is the off-diagonal Green's function describing the electromagnetic interaction between two dipole moments $\mu_{1,2}$ of frequency $\omega_0$. The rates of diagonal and off-diagonal dissipations are given by
\begin{equation}
\label{decay}
\gamma_{ij}=\frac{2 \omega_0^2}{\epsilon_0 c^2 \hbar}Im[\mu_i^{*} {\bf G}(\omega_0,{\bf r}_i,{\bf r}_j)\mu_j],
\end{equation}
with $i,j=1,2$, $\gamma_{12}=\gamma_{21}$ and we will label $\gamma_{ii}=\gamma_i$.

It was recently found\cite{martincano10a,dzsotjan11a} that when the propagating
plasmon supported by the PW is the dominant channel
for emission (i.e., large $\beta$ factor), a very good approximation for the total Green's function can be obtained by
only including its plasmon contribution, $\bf{G}(\bf{r}_1,\bf{r}_2)\approx \bf{G_p}(\bf{r}_1,\bf{r}_2)$. In this way, analytical expressions for both $g_{12}$ and $\gamma_{12}$ can be easily derived:
\begin{eqnarray}
\label{coupling-crossdecay}
g_{12} & = & \frac{\gamma}{2} \beta e^{-d/(2 L)} \sin(k_{pl} d ) \nonumber \\
\gamma_{12} & = & \gamma \beta e^{-d/(2 L)} \cos(k_{pl} d),
\end{eqnarray}

where $d$ is the separation between the two qubits, $k_p$ and $L$ being the wave-number and propagation length of the plasmon, respectively. The crucial point of Eq. \ref{coupling-crossdecay}
is that the $\pi/2$ phase shift between the coherent and
incoherent parts of the coupling mediated by plasmons
enables to switch off one of the two contributions while
maximizing the other by just tailoring the inter-qubit
distance, opening the possibility of modulating the degree
of entanglement as we will see in the next sections.

\section{Spontaneous decay.\label{sec:spontaneous}}

In order to solve the dynamics given by the master equation\ref{masterequation} we need to project in some basis this equation and then solve the set of differential equations for different initial conditions depending on our initial state. A possible basis to represent the dynamics (\ref{masterequation}) is the one given by \{$\ket{0}=\ket{g_1,g_2},\ket{1}=\ket{e_1,g_2},\ket{2}=\ket{g_1,e_2},\ket{3}=\ket{e_1,e_2}$\}, where $g_i/e_i$ labels the ground/excited state of the $i$-qubit. However, for a better understanding of the physics behind entanglement it is convenient to change to the basis
\{$\ket{0}=\ket{g_1,g_2},\ket{\pm}=\frac{1}{\sqrt{2}}(\ket{1}\pm\ket{2}),\ket{3}=\ket{e_1,e_2}$\}, where two maximally entangled states $\ket{\pm}$ appear explicitely. This basis also allows to understand more clearly the dynamics of the system as diagonalize the coherent part from the dynamics. In figure \ref{fig1} (b) could be observed that depending on both the sign and absolute value of the cross decay term $\gamma_{12}$ one of the states $\ket{\pm}$ could be decoupled from the dynamics of the rest of states.

Once the density matrix $\rho (t)$ is obtained by numerically solving Eq. (\ref{masterequation}) represented in the mentioned basis, the entanglement of the two qubits is quantified by means of the concurrence~\cite{wootters98a}, $C\equiv[\mathrm{max}\{0,\sqrt{\lambda_1}-
\sqrt{\lambda_2}-\sqrt{\lambda_3}- \sqrt{\lambda_4}\}]$, where
$\{\lambda_1, \lambda_2, \lambda_3, \lambda_4 \}$ are the eigenvalues
in decreasing order of the matrix $\rho \mathrm{T} \rho^*\mathrm{T}$, with $\mathrm{T}$ being the
anti diagonal matrix with elements $\{ -1,1,1,-1\}$. The two main ingredients of our problem affect $C$ in a different way. The coherent coupling $g_{12}$ produces oscillations, whereas the effect from the cross-decay $\gamma_{12}$ is much more important non-oscillatory contribution to $C$.

Now we will describe two situations where the system is initialized in a given state  from which it decays spontaneously. In the first one, the system is initially prepared in a one-excitation unentangled state and in the second one a doubly-excited state.

\subsection{Spontaneous formation of entanglement.\label{subsec:eof}}

\begin{figure}
\begin{center}
\includegraphics[width=0.99\linewidth,angle=0]{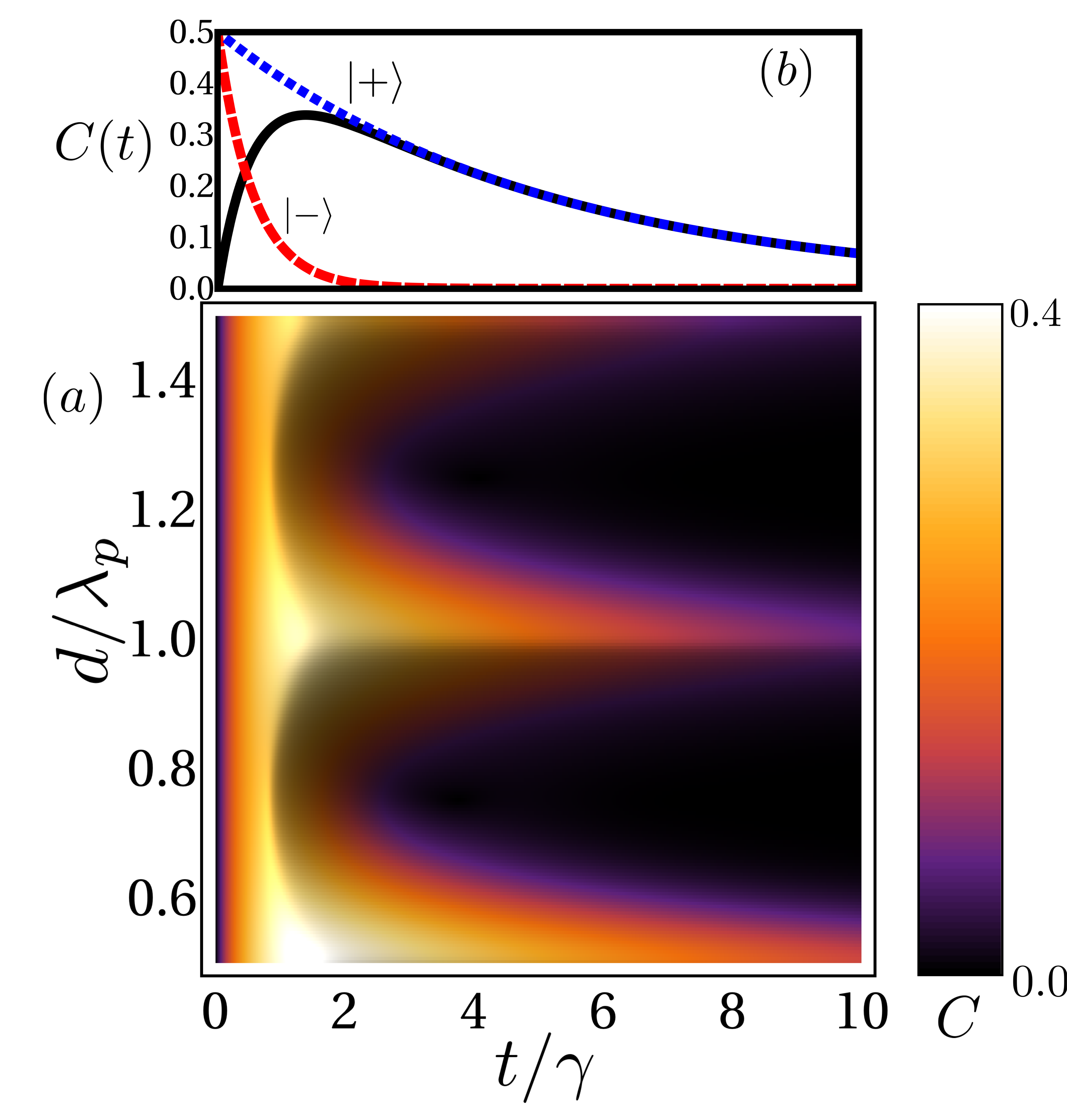}
\end{center}
\caption{(Color online) Panel (a) Density plot of concurrence as a function of time and distance between qubits when the system is initially prepared at the state $\ket{1}$. It corresponds to two qubits entangled by means of a SPP waveguide with $\beta e^{-d/(2 L)}=0.94$, with $\lambda_p=640 nm$ and $\gamma_1=\gamma_2=\gamma$.The spontaneous formation of entanglement is observed. Panel (b) show the concurrence dynamics for $d=\lambda_{pl}$ in black and the $\ket{+}/\ket{-}$ population in dotted blue/dashed red line respectively.}
\label{spontaneous1}
\end{figure}

In order to analyze the spontaneous formation of entanglement in the system, one can initially prepare the system, for instance in the state $\ket{1}$, by a $\pi$-pulse on one of the qubits. Then the only non-zero elements of the initial density matrix are: $\rho_{++}(0)=\rho_{--}(0)=\rho_{+-}(0)=\rho_{-+}(0)=1/2$, so the dynamics is confined to the subspace spanned by these three vectors: \{$\ket{0},\ket{\pm}$\}. Apart from the populations, the only non-zero elements in the dynamics are $\rho_{\pm\mp}(t)$. The expressions for the non-zero density matrix elements in this case are given by:

\begin{eqnarray}
 \label{rateequations1}
\dot{\rho}_{++}(t)=-\gamma_{+}(\rho_{++}(t))\nonumber\\
\dot{\rho}_{--}(t)=-\gamma_{-}(\rho_{--}(t))\nonumber\\
\dot{\rho}_{-+}(t)=-(\gamma-2 i g_{12})\rho_{-+}(t)
\end{eqnarray}
where $\gamma_{\pm}=\frac{\gamma_1+\gamma_2}{2}\pm\gamma_{12}$. In this case, the concurrence takes the form:
\begin{eqnarray}
C(t)\sqrt{[\rho_{++}(t)-\rho_{--}(t)]^2+4\Im [\rho_{+-}(t)]^2}
\label{c(t)}
\end{eqnarray}

and solving Eq. \ref{rateequations1} we arrive to the following simple expression for concurrence:
\begin{equation}
 \label{conc1}
C(t)=e^{-\gamma t} \sqrt{\sinh^2(\gamma_{12} t)+\sin(2 g_{12} t)}
\end{equation}

where we can see there are two different contributions: the one from the coherent coupling $g_{12}$ which is responsible for the oscillations, and a non-oscillatory contribution corresponding to $\gamma_{12}$ which produces an asymmetric population of the $\ket{\pm}$ states responsible for the lengthening of the lifetime of the entanglement in this particular situation. The dynamics of $C$ and its dependence with the inter-qubit distance $d$ is shown in figure \ref{spontaneous1} for a  PW with $\beta e^{-d/(2 L)}=0.94$. For all $d$ one could observe an fast spontaneous formation of entanglement, however for $d=n\lambda_p/2$ with $n=1,2,...$ its decay is much slower. This is mainly due to the large asymmetry in the timescale of the two cascades in Fig. \ref{fig1} which produces a large imbalance of populations  and therefore creating entanglement as predicted in Eq. \ref{c(t)}.

\subsection{Sudden birth and revival of entanglement.\label{subsec:birthrevival}}

Since one of the goals of this paper is to show all the entanglement phenomenology of our particular system we are going to focus now in a different  situation where our  system is initially prepared in a non-maximally entangled state: $\ket{\Phi_0}=\sqrt{q}\ket{0}+\sqrt{1-q}\ket{3}$, where $0\leq q \leq 1$. Consequently the initial non-zero elements of the density matrix are: $\rho_{00}(0)=q,\rho_{33}(0)=1-q,\rho_{03}=\sqrt{q(1-q)}$. So, as in the previous case, the rest of the density matrix will remain zero except for the symmetric and antisymmetric populations that will build up during the evolution. The expressions for the non-zero elements are given by:
\begin{eqnarray}
 \label{revivalquations11}
\rho_{33}(t)=q e^{-2\gamma t} \nonumber \\
\rho_{++}(t)=q\frac{\gamma_+}{\gamma_-} e^{-2\gamma t}\Big[e^{\gamma_{-}t}-1 \Big]\nonumber\\
\rho_{--}(t)=q\frac{\gamma_-}{\gamma_+} e^{-2\gamma t}\Big[e^{\gamma_{+}t}-1 \Big]\nonumber\\
\rho_{03}(t)=\sqrt{q(1-q)} e^{-(\gamma-2 i \omega_0) t}
\end{eqnarray}

After a tedious but simple algebra you arrive to the following expression for the concurrence $C(t)=\mathrm{max}\{0,C_1(t),C_2(t)\}$, where:
\begin{eqnarray}
\label{concure2}
C_1(t)=2|\rho_{03}(t)|-|\rho_{++}(t)+\rho_{--}(t)| \nonumber\\
C_2(t)=-2\sqrt{\rho_{00}(t)\rho_{33}(t)}+|\rho_{++}(t)-\rho_{--}(t)|
\end{eqnarray}

\begin{figure}
\begin{center}
\includegraphics[width=0.99\linewidth,angle=0]{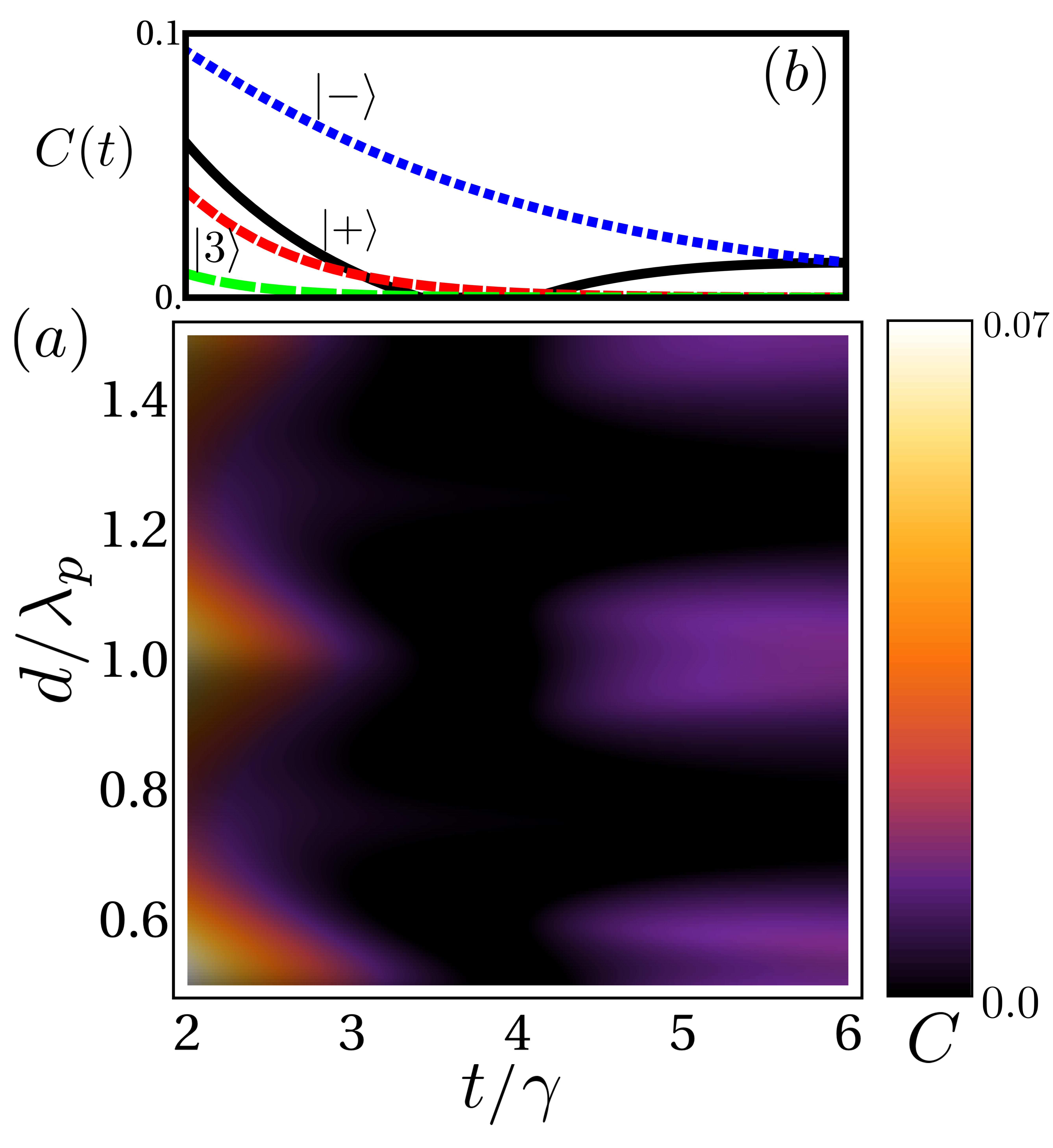}
\end{center}
\caption{(Color online)Panel (a) Density plot of concurrence as a function of time and distance between qubits when the system is initially prepared at the state $\ket{\Phi_0}$ defined as in the text with $q=1/2$. It corresponds to two qubits entangled by means of a SPP waveguide with $\beta e^{-d/(2 L)}=0.94$, with $\lambda_p=640 nm$ and $\gamma_1=\gamma_2=\gamma$. The so-called revival of entanglement is observed. Panel (b) show the concurrence dynamics for $d=\lambda_{pl}$ in black and the $\ket{+}/\ket{-}/\ket{3}$ population in dotted blue/dashed red/long-dashed green line respectively.}
\label{spontaneousrevival}
\end{figure}

This double criterion for entanglement leads to a very interesting and well-known phenomena which has been called as \textit{revival} of entanglement\cite{ficek06a}. In figure \ref{spontaneousrevival} we have plotted the concurrence for a system with $q=0.5$, which is initially entangled. It could be seen that after a short time, of the order $t/\gamma=4$ entanglement disappears for a time, but latter it, counterintuitively, revives. The entanglement revival is clearly related with $\gamma_{12}$ as we could see that in the distances where it is small, the revival does not take place. Counterintuitively, as the  $\{\ket{1},\ket{2}\}$ (or $\ket{\pm}$) are initially uncoupled, and remain uncoupled forever, the coherent coupling $g_{12}$ of the dipole does not play a significant role in the dynamics of the system, as it can only change one excitation, but not two. Quantitatively the revival is very subtle as it only gets a maximum entanglement of around $3 \%$.

Considering the limiting case where $q=1$, so that our initial state is given by $\ket{\Phi_0}=\ket{3}$, which is unentangled, then the only non-zero element of the initial density matrix is given by $\rho_{33}=1$. Therefore the $\rho(t)$ remains diagonal during its evolution, being the populations the only non-zero elements:
\begin{eqnarray}
 \label{revivalquations111}
\rho_{33}(t)= e^{-2\gamma t} \nonumber \\
\rho_{++}(t)=\frac{\gamma_+}{\gamma_-} e^{-2\gamma t}\Big[e^{\gamma_{-}t}-1 \Big]\nonumber\\
\rho_{--}(t)=\frac{\gamma_-}{\gamma_+} e^{-2\gamma t}\Big[e^{\gamma_{+}t}-1 \Big]\nonumber\\
\end{eqnarray}

One would expect then that no entanglement could build up from this initial state, however, after some simple algebra you arrive to the following expression for the concurrence $C(t)=\mathrm{max}\{0,C_2(t)\}$ where:

\begin{eqnarray}
 \label{concure3}
C_2(t)=|\rho_{++}(t)-\rho_{--}(t)|-2\sqrt{\rho_{00}(t)\rho_{33}(t)}
\end{eqnarray}

\begin{figure}
\begin{center}
\includegraphics[width=0.99\linewidth,angle=0]{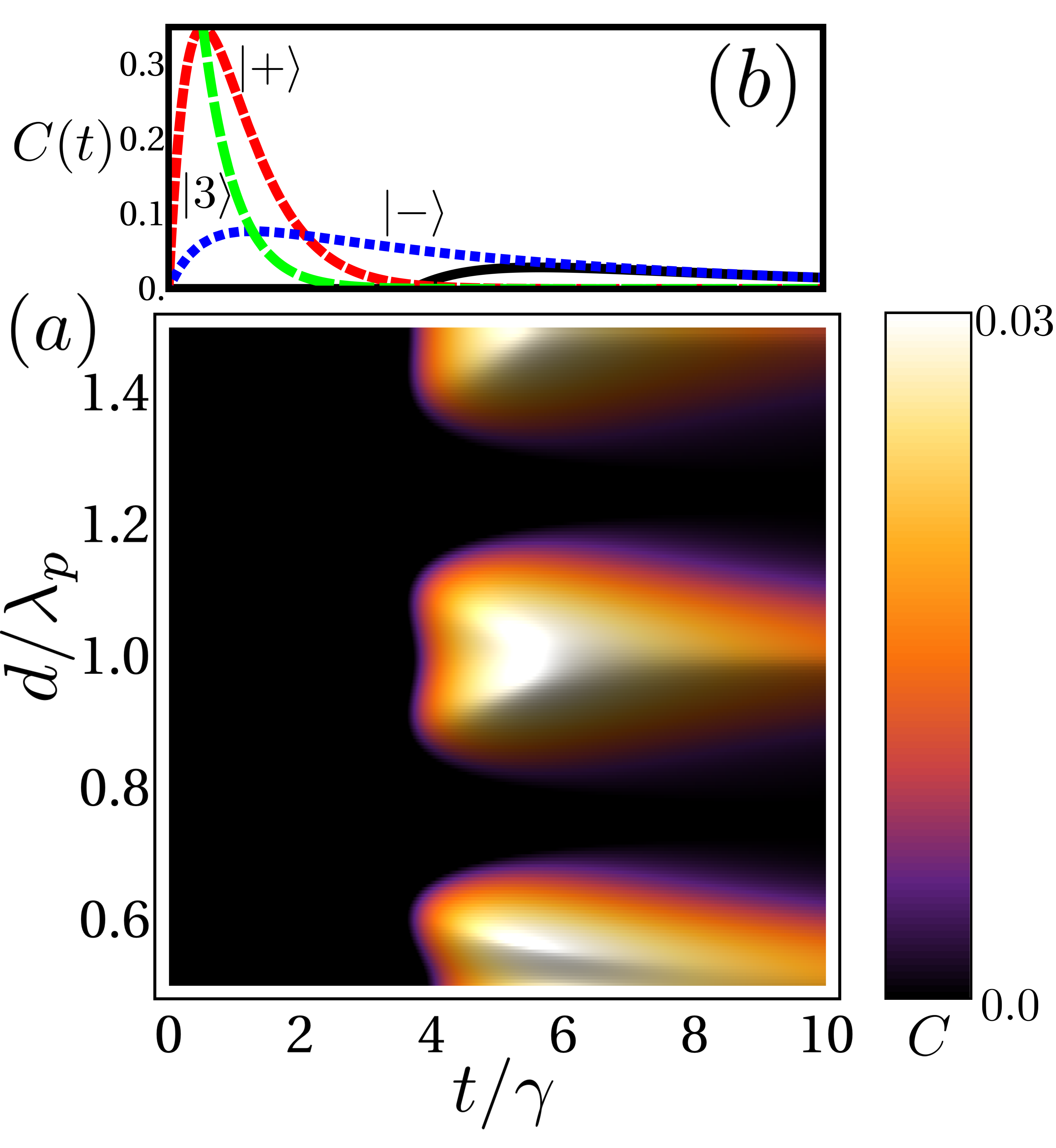}
\end{center}
\caption{(Color online)Panel (a): Density plot of concurrence as a function of time and distance between qubits when the system is initially prepared in the state $\ket{3}$. It corresponds to two qubits entangled by means of a SPP waveguide with $\beta e^{-d/(2 L)}=0.94$, with $\lambda_p=640 nm$ and $\gamma_1=\gamma_2=\gamma$. The so-called sudden-birth of entanglement is observed. Panel (b) show the concurrence dynamics for $d=\lambda_{pl}$ in black and the $\ket{+}/\ket{-}/\ket{3}$ population in dotted blue/dashed red/long-dashed green line respectively}
\label{spontaneousbirth}
\end{figure}

So if the difference of population between the symmetric and antisymmetric states overcomes the second term in Eq. \ref{concure3}, some concurrence may build up. Again, the crucial element is the collective decay rate $\gamma_{12}$ and not the coherent coupling $g_{12}$, which does not play a role. In figure \ref{spontaneousbirth} this phenomenon could be observed for a given set of parameters(cf. caption): no concurrence occurs for the initial times and suddenly a $2\%$ is built. This phenomenon usually called \textit{sudden birth}\cite{lopez08a,ficek08a} of entanglement, is again subtle in our system but still observable.

\section{Steady-state entanglement.\label{sec:steady}}

Up to now, we have seen that it is possible to observe finite-time entanglement in different situations. Nevertheless, one is usually interested in having a stationary state with some degree of entanglement. In order to achieve a stationary-state a continuous pumping is required. The pumping of the qubits may have incoherent nature\cite{delValle11a} or one could use coherent excitation, i.e. a laser field. We are going to assume a situation for sufficiently separated qubits, where the stationary state can be modulated by acting  independently on each qubit with a laser of Rabi frequency $\Omega_i$. Therefore a new term, $\sum_i \hbar \Omega_i(\sigma^\dagger_i+\sigma_i)$ must be included in the coherent dynamics, i.e. in Eq.(\ref{hamilt}).

\begin{figure}
\begin{center}
\includegraphics[width=0.99\linewidth,angle=0]{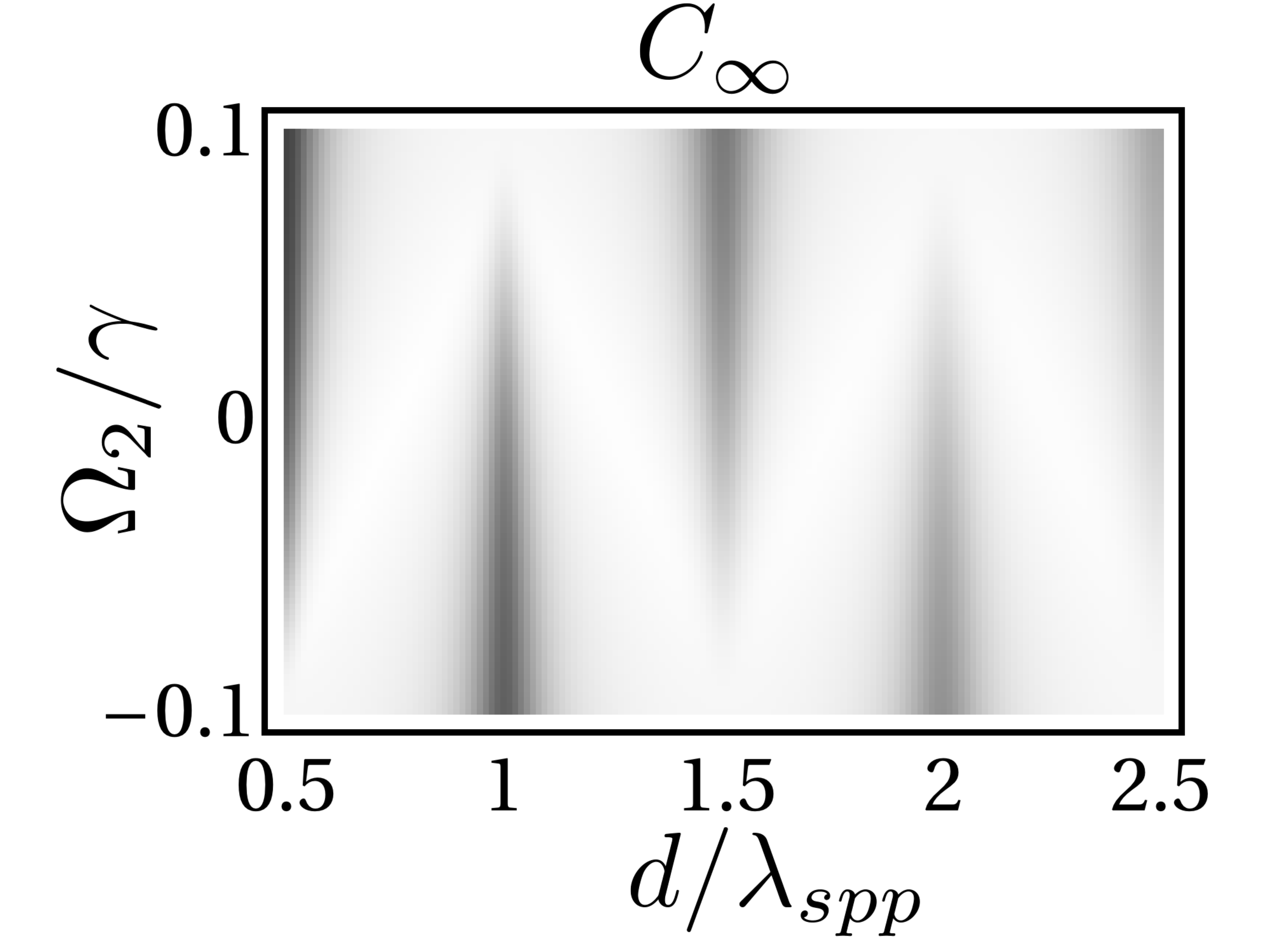}
\end{center}
\caption{Steady state concurrence as a function of the separation between two equal qubits for $\beta=0.94$, $L=2 \, \mu$m and for a pumping configuration with $\Omega_1=0.1\gamma$ and $\Omega_2$ ranging from $-\Omega_1$ to $\Omega_1$. Black regions correspond to higher values of concurrence while the white regions correspond to zero concurrence. }
\label{fig:Concsteady}
\end{figure}

As we recently showed\cite{gonzaleztudela11a} one could find steady-state entangled situations playing with the laser intensities. In Fig. \ref{fig:Concsteady}, we have plotted the steady-state concurrence for the typical set of parameters that we have been considering(cf. caption) within the paper as a function of the inter-qubit distance and for different pumping configurations. In this figure, the pumping on the first qubit is fixed: $\Omega_1=0.1\gamma$, while the second laser intensity $\Omega_2$ ranges from $-\Omega_1$ to $\Omega_1$, going then from an antisymmetric pumping with $\Omega_1=-\Omega_2$ where one can observe that peaks in the concurrence appear for $d$ close to an even multiples of  $\lambda_{pl}/2$. As expected, in the symmetric pumping situation ($\Omega_1=\Omega_2$) the peaks appear for odd multiples of $\lambda_{pl}/2$. In the intermediate region of asymmetric pumping, $\Omega_1\neq \Omega_2$, the periodicity is broken and both even and odd multiples of $\lambda_{pl}/2$ appear.

This possibility of modulating the steady-state of a system by changing the pumping and assisted by dissipation is connected with the ideas of dissipative computation and state-engineering recently proposed\cite{verstraete09a} and experimentally demonstrated\cite{krauter10a}.

\subsection{Entanglement-purity diagram.\label{subsec:steady}}

Up to know we have only worried about the quantification of the degree of entanglement using the Wooters concurrence. However, as we are dealing with a dissipative environment, the quantum state of the system is never pure and its degree of purity is very relevant for its possible application  in quantum information protocols\cite{bose00b} As we have already seen, we use a density matrix description which is a very powerful formalism to describe \textit{mixed states}. One standard measurement for characterizing the degree of mixture of a system with a given $\rho$ density matrix is the \textit{linear entropy}\cite{bose00a} defined by:
\begin{equation}
 \label{linearentropy}
S_L=\frac{4}{3}\big(1-\Tr(\rho^2)\big)
\end{equation}

\begin{figure}
\begin{center}
\includegraphics[width=0.99\linewidth,angle=0]{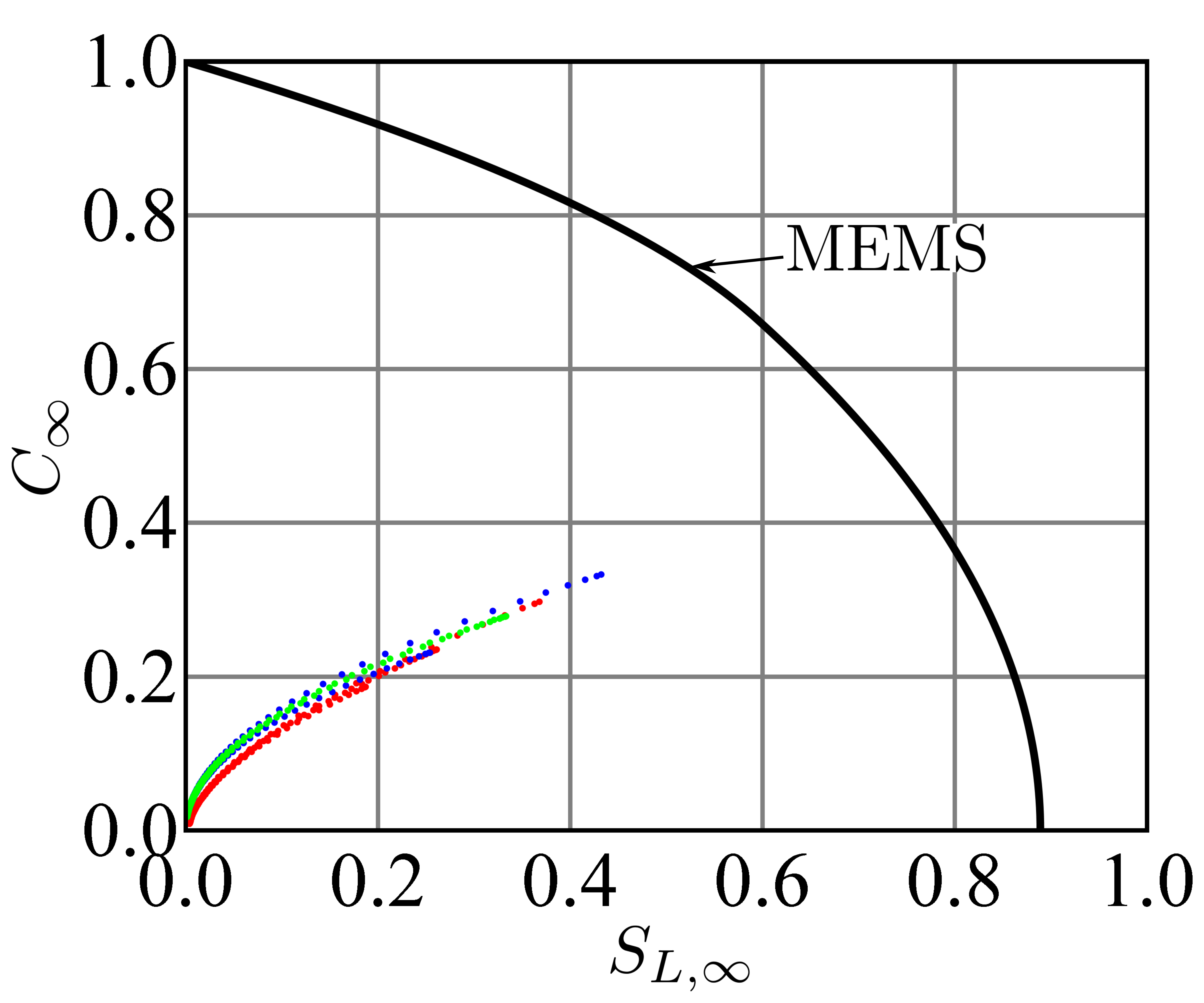}
\end{center}
\caption{(Color online) Concurrence-Linear entropy diagram for a system with parameters $\beta=0.94$, $L=2 \, \mu$m with each point corresponding to different inter-qubit distance ranging from $d/\lambda_{spp}=0.5-1.5$, whereas the three different color correspond to three different pumping configurations: $\Omega_1=0.15\gamma$ and $\Omega_2=0$ (red), $\Omega_1=\Omega_2=0.1\gamma$ (blue), and $\Omega_1=-\Omega_2=0.1\gamma$ (green)}
\label{fig:ConcEntropy}
\end{figure}

When one has a pure state then $\Tr(\rho^2)=1$ so the linear entropy becomes $S_L=0$ whereas for a maximally mixed state  $\Tr(\rho^2)=1/4$ so that $S_L=1$. However, the most interesting regime for us is what happens in between when some entanglement is present but the purity of your system is not perfect. It is well-known that if the density matrix describing a two-qubit system with a certain degree of mixture, then there is a maximum degree of entanglement this system may achieve by means of unitary transformations. These states are usually called \textit{Maximally Entangled Mixed States} (MEMS) and were first proposed by Ishizaka and Hiroshima\cite{Ishizaka00a} and they occupy a region the concurrence-linear entropy diagram\cite{munro01a} shown in black in Fig. \ref{fig:ConcEntropy}. We have also included the points resulting from the calculations in our system for the three different pumping configuration that we aforementioned in the previous section (see the caption for details) so that one could see how far from the optimal entanglement is our state for a certain degree of purity.

%
%
%
%

\section{Conclusions.\label{sec:conc}}

We have used the properties of the qubit-qubit coupling induced by plasmon-polariton nanowaveguides\cite{martincano10a,gonzaleztudela11a} to explore a wide range of entanglement situations: starting from unentangled states we have shown the possibility of spontaneous formation, sudden birth, and revival of entanglement. Using a pumped configuration scheme, we have been able to achieve steady-state modulable entanglement and characterize the quantum state in the concurrence-entropy diagram, comparing them to the MEMS. The possibility of modulating entanglement open the possibility of implementing the ideas of dissipative quantum computation\cite{verstraete09a} in this type of plasmonic systems.

\section*{Acknowledgements.\label{sec:ack}}

Work supported by the
Spanish MICINN (MAT2008-01555, MAT2009-06609-C02, CSD2006-00019-QOIT and CSD2007-046-
NanoLight.es) and CAM
(S-2009/ESP-1503). A.G.-T. and D.M.-C acknowledge FPU grants (AP2008-00101 and AP2007-00891, respectively) from the Spanish Ministry of Education.

\bibliography{plasmon1}

\end{document}